%% file: main.tex
\documentclass[sigconf, screen]{acmart}
\usepackage[nolist]{acronym}
\usepackage{subcaption}
\usepackage{xspace}
\usepackage{gensymb}
\usepackage[acronym]{glossaries}
\makeglossaries

\AtBeginDocument{%
  }

\copyrightyear{2026}
\acmYear{2026}
\setcopyright{cc}
\setcctype{by}
\acmConference[CHI EA '26]{Extended Abstracts of the 2026 CHI Conference on Human Factors in Computing Systems}{April 13--17, 2026}{Barcelona, Spain}
\acmBooktitle{Extended Abstracts of the 2026 CHI Conference on Human Factors in Computing Systems (CHI EA '26), April 13--17, 2026, Barcelona, Spain}
\acmDOI{10.1145/3772363.3799273}
\acmISBN{979-8-4007-2281-3/2026/04}



\begin{document}

\title{\textsc{MRDrive}: An Open Source Mixed Reality Driving Simulator for Automotive User Research}


\author{Patrick Ebel}
\email{ebel@uni-leipzig.de}
\orcid{0000-0002-4437-2821}
\affiliation{%
  \institution{ScaDS.AI, Leipzig University}
  \city{Leipzig}
  \country{Germany}}

\author{Michał Patryk Miazga}
\email{miazga@uni-leipzig.de}
\orcid{0009-0003-5579-3036}
\affiliation{%
  \institution{ScaDS.AI, Leipzig University}
  \city{Leipzig}
  \country{Germany}}

\author{Martin Lorenz}
\email{martin.lorenz@uni-leipzig.de}
\orcid{0009-0003-3517-5839}
\affiliation{%
  \institution{ScaDS.AI, Leipzig University}
  \city{Leipzig}
  \country{Germany}}

\author{Timur Getselev}
\email{tg12lera@studserv.uni-leipzig.de}
\orcid{0009-0001-2116-6590}
\affiliation{%
  \institution{Leipzig University}
  \city{Leipzig}
  \country{Germany}}

\author{Pavlo Bazilinskyy}
\email{p.bazilinskyy@tue.nl}
\orcid{0000-0001-9565-8240}
\affiliation{%
  \institution{Eindhoven University of Technology}
  \city{Eindhoven}
  \country{The Netherlands}}

\author{Celine Conzen}
\email{cz67wuse@studserv.uni-leipzig.de}
\orcid{0009-0004-2835-9848}
\affiliation{%
  \institution{ScaDS.AI, Leipzig University}
  \city{Leipzig}
  \country{Germany}}

%%
%% By default, the full list of authors will be used on the page
\renewcommand{\shortauthors}{Ebel et al.}
\newcommand{\name}{\textsc{MRDrive}\xspace}

\begin{abstract}
Designing and evaluating in-vehicle interfaces requires experimental platforms that combine ecological validity with experimental control. Driving simulators are widely used for this purpose. However, they face a fundamental trade-off: high-fidelity physical simulators are costly and difficult to adapt, while virtual reality simulators provide flexibility at the expense of physical interaction with the vehicle. In this work, we present \name, an open mixed-reality driving simulator designed to support HCI research on in-vehicle interaction, attention, and explainability in manual and automated driving contexts. \name enables drivers and passengers to interact with a real vehicle cabin while being fully immersed in a virtual driving environment. We demonstrate the capabilities of \name through a small pilot study that illustrates how the simulator can be used to collect and analyze eye-tracking and touch interaction data in an automated driving scenario. \name is available at: \url{https://github.com/ciao-group/mrdrive}
\end{abstract}

\begin{teaserfigure}
    \centering
  \includegraphics[width=\textwidth]{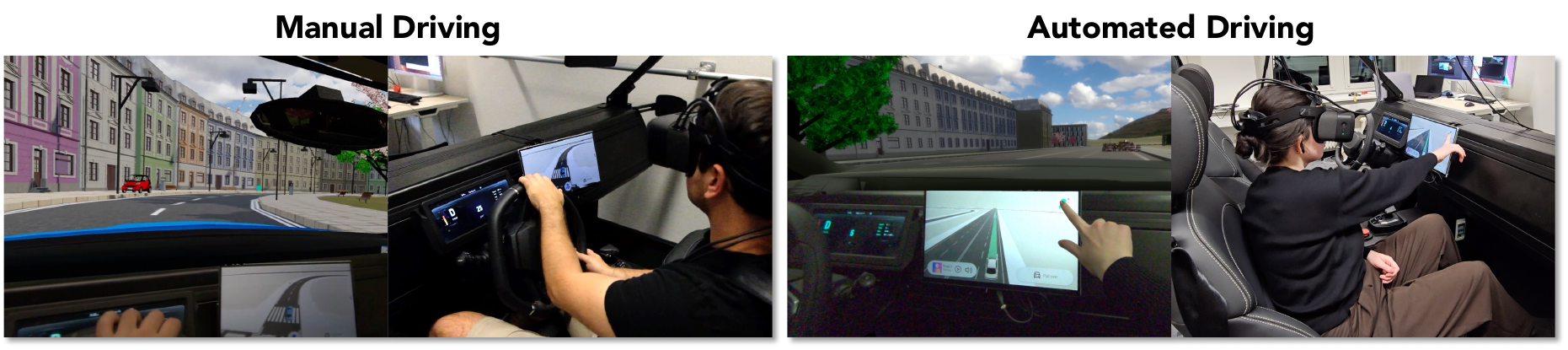}
  \caption{\name enables drivers and passengers of manual and automated vehicles to be immersed in a 360\degree virtual driving environment while the physical in-cabin environment is rendered via pass-through, allowing for natural interaction with the vehicle.}
  \Description{TBD}
  \label{fig:teaser}
\end{teaserfigure}


\include{acronyms}

\maketitle

\section{Introduction}

%Current Driving Simulators
Designing and evaluating in-vehicle interfaces requires experimental platforms that combine ecological validity with experimental control. Driving simulators~\cite{fisher_handbook_2011} are widely used for this purpose, as they enable safe, repeatable studies of driver behavior and interaction~\cite{caird2011twelve}. 
Such simulators are intended to support inferences about real-world driving, which in turn requires that they provide a sufficiently detailed replication of the driving task and its context.
However, simulators that go beyond low-fidelity “screen-and-steering-wheel” setups are typically either closed-source~\cite{mortimer_hector-vr_2019, krueger_silabtask_2005} or tightly coupled to proprietary and costly projection- and motion-based hardware~\cite{himmels_bigger_2024}, or both. At the opposite end of the spectrum, low-cost simulators sacrifice physical realism, limiting the ecological validity of interaction behavior~\cite{mcmahan_evaluating_2012}.

%VR Driving Simulators as a potential solution
An opportunity to increase immersion without relying on expensive and space-intensive projection systems lies in \gls{VR} driving simulators. VR-based simulators provide compact, comparatively low-cost, yet immersive evaluation environments~\cite{bayarri_virtual_1996}. Recent systems are increasingly built on free-to-use game engines such as Unity~\cite{bazilinskyy_coupled_2020}, improving accessibility beyond large, well-funded research labs.
Beyond cost and space efficiency, VR simulators offer several methodological advantages for studying in-vehicle interaction. They support a full $360^{\circ}$ rendering of the driving environment, enabling high perceptual immersion, and modern \glspl{HMD} integrates eye, head, and hand tracking directly into the \gls{HMD}, allowing rich behavioral data to be captured without extensive external instrumentation. VR simulations also enable spatial and three-dimensional interface elements to be embedded directly into the task environment, supporting the exploration of novel interaction concepts, including gesture-based interaction.
However, a key limitation of VR driving simulators for evaluating in-vehicle interfaces is the loss of physical interaction with the vehicle interior and other devices such as infotainment systems or smartphones, which have been shown to affect driving behavior and safety~\cite{ebel_forces_2023, ebel_multitasking_2023}. The absence of a high-fidelity vehicle cabin and realistic haptic feedback raises concerns about ecological validity. Virtual touch interactions often require users to adapt their motor actions to abstract controllers instead of directly interacting with physical surfaces, and the lack of realistic tactile and material engagement can distort task timing and precision~\cite{lavoie_how_2025, personeni_ecological_2023}.

We argue that \gls{MR}~\cite{rauschnabel_what_2022} driving simulators, where the driving environment is ``virtual'' and the interior is ``real'' combine the best of both worlds. They enables $360^{\circ}$ immersion in the task environment without requiring the space of a projection-based simulator while retaining realistic in-vehicle interaction with the driving controls, infotainment system, or potential passengers.

Accordingly, we introduce \name, an open-source Mixed-Reality Driving Simulator designed to support a broad spectrum of AutoUI research. \name is based on Unity, provides various driving scenes, a modern infotainment system, and template scenarios. It supports manual and automated driving and, due to its modular architecture, is easily extensible. 
With \name, our goal is to lower the barrier to entry for high-fidelity MR-based driving studies and to enable reproducible experimentation across labs. To support the open science movement in the automotive user research community~\cite{ebel_changing_2024}, \name is fully open source and can be adjusted to local hardware requirements.
We demonstrate the capabilities of \name through a small pilot study that illustrates how the simulator can be used to collect and analyze eye-tracking and touch interaction data in an automated driving scenario.
\textbf{\name makes the following contributions: }

\begin{itemize}
    \item A Unity-based open-source \gls{MR} driving simulator.
    \item An extensible in-vehicle infotainment system inspired by modern \glspl{AV}.
    \item Example scenarios, showcasing how to run user studies and collect driving data, interaction data, and physiological measurements.
\end{itemize}

\section{Related Work}
% \cite{himmels_bigger_2024}: While there were fewer differences in driving behavior and in distance and speed perception than expected, there were clear benefits of higher fidelity in the context of simulator sickness.
%  \cite{yeo_toward_2020}: comparison of different levels of immersiveness in driving simulation

Recently, various XR driving simulators have been developed. However, these simulators vary not only in how they leverage XR technologies but also in whether they open-source their software, thereby making it accessible to the research community.

An example of an open-source driving simulator is provided by \citet{bazilinskyy_coupled_2020}. Their \textit{Coupled Simulator} targets research on interactions between pedestrians and (co-) drivers in both manual and \glspl{AV}. It supports a range of input devices, including head-mounted displays, motion suits, and game controllers, allowing participants to assume the role of either vehicle occupants or pedestrians. Built on the Unity game engine, the simulator can be operated in both \gls{VR} and conventional screen-based configurations. The simulator has been used in multiple coupled human–human interaction studies, including investigations of driver–pedestrian encounters~\cite{bazilinskyy2022get,mok2022stopping}.

With \textit{Hector-VR}, \citet{mortimer_hector-vr_2019} presents a \gls{VR} driving simulator designed to evaluate driving and in-vehicle interaction behavior in older adults. \textit{Hector-VR} combines a physical vehicle with an Oculus Rift \gls{HMD}. Although the vehicle provides a physically realistic cockpit, participants experience the driving environment entirely in \gls{VR}. The simulator software is proprietary and tightly coupled to the hardware. It follows a systems-of-systems architecture intended to support modularity as well as straightforward installation and maintenance.
Another approach to \gls{MR} driving simulation is presented by projects like \textit{VR-OOM}~\cite{goedicke_vr-oom_2018} and \textit{MAXIM}~\cite{yeo_maxim_2019}. Here, participants wear an \gls{HMD} and are fully immersed in a \gls{VR} while sitting in a real vehicle that is driven by a test driver. This \textit{Wizard-of-Oz} setup combines the benefits of real-world driving dynamics with full experimental control over the visually perceived driving environment. While \textit{VR-OOM} is openly available, \textit{MAXIM} is closed-source. While experimental results are promising, these simulators require a real vehicle, road infrastructure, and participants do not experience the physical car interior. 
With \textit{XR-OOM}, \citet{goedicke_xr-oom_2022} present an open-source MR driving simulation in which the driver operates a real car while wearing an \gls{HMD} that superimposes virtual objects on the road. The driver experiences the real driving environment via pass-through vision. Thus, XR-OOM offers the opportunity to test and develop future interfaces in quasi-real-world environments, though it still requires a real vehicle and test track.
One approach similar to \name is the \textit{UniNet} driving simulator presented by \citet{arppe_uninet_2020}. This approach is based on the Unity game engine and involves participants sitting in a green screen chamber with only a steering wheel. Through chroma keying, the videos from the passthrough cameras in the HMD and the driving simulation are composed in a way that allows participants to see their arms and the physical steering wheel while the rest of the environment is rendered digitally. Although no physical motion can be replicated, this work is the first to partly blend real in-cabin artifacts with a simulated driving environment.
Together, existing XR driving simulators trade off openness, physical fidelity, and experimental control in different ways.

\name occupies a gap in the current ecosystems of driving simulators by combining open-source software with a simulation environment that supports various levels of cabin fidelity, enabling reproducible \gls{MR}-based prototyping without relying on on-road tests or proprietary infrastructure.

\begin{figure*}
    \centering
  \includegraphics[width=0.9\textwidth]{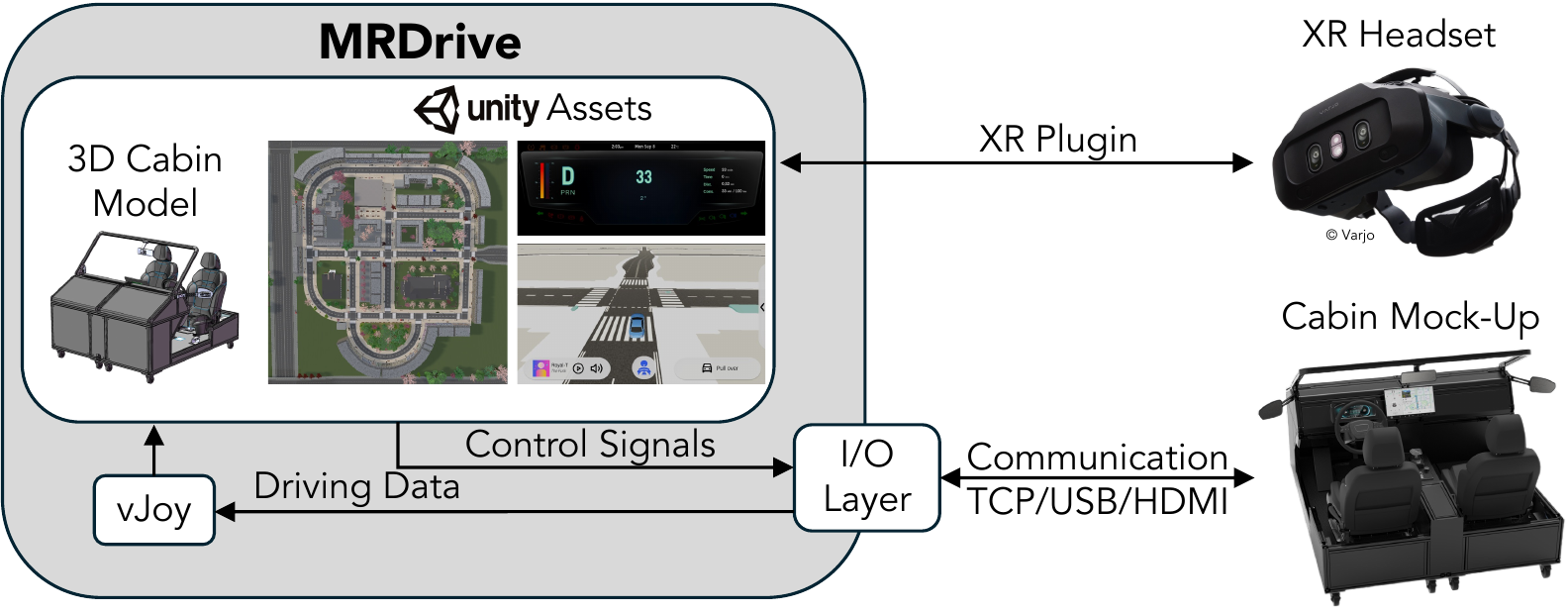}
  \caption{Overview of the \name system architecture.}
  \Description{The figure illustrates the main software components of \name and their interactions with the underlying hardware. It highlights the flow of communication between modules and how data and control signals are exchanged across the system.}
  \label{fig:workflow}
\end{figure*}

\section{\name}

\name is an open source mixed reality driving simulator designed to support HCI research on in-vehicle interaction, attention, and explainability in manual and automated driving contexts. 

\subsection{Software}
The architecture of \name, illustrated in \autoref{fig:workflow}, comprises three primary modules. 
\name integrates a Unity-based driving environment and orchestrates the data exchange between the cabin mock-up and the \gls{HMD}. 
\name receives steering and vehicle control inputs from the mock-up via TCP and uses vJoy\footnote{\url{https://sourceforge.net/projects/vjoystick/}} to map these inputs to corresponding control signals within the Unity environment.
\name further receives touch input events from the \gls{IVIS} via USB.
In addition, \name connects to the \gls{HMD} via the Unity XR plugin and receives positioning information as well as eye-tracking information. 
\name integrates all signals in the Unity simulation logic to continuously update vehicle dynamics, traffic behavior, and the state of interactive user interfaces. 
Based on the updated simulation state, \name renders both the immersive virtual environment for the \gls{HMD} and the in-cabin user interfaces, such as the instrument cluster and center stack touchscreen.
It then transmits force feedback signals to the mock-up via TCP, outputs the rendered user interfaces to the mock-up via HDMI, and the rendered virtual driving scene to the \gls{HMD}

\paragraph{\textbf{IVIS}}
The \gls{IVIS} is implemented as two Unity prefabs, the instrument cluster and the center stack touchscreen, and provides interactive, realistic user feedback. The instrument cluster displays core driving information (e.g., speed, gear, and control inputs) in real time, along with additional vehicle metrics. To mimic a modern vehicle dashboard, all interface elements briefly illuminate during startup.
The design of the center stack touchscreen is inspired by those of modern automated cars, such as Waymo\footnote {\url{https://waymo.com/}}. It shows the current position of the ego-vehicle and displays the map in a bird's-eye view. Objects in the surroundings are shown as contours, indicating that the automated driving software has detected them. The touchscreen also offers a welcome screen and music playback with volume control. Both IVIS assets can be projected onto the touchscreen for direct user interaction and modification.

\paragraph{\textbf{Driving Environment}}
For the driving environment, we expanded upon the Coupled-Simulator platform developed in Unity by \citet{bazilinskyy_coupled_2020}. We selected this platform because of its open-source architecture and modular design. We reused the street network and car and pedestrian prefabs. We replaced all other assets and added new ones, including the IVIS and a vehicle cabin mock-up. We also reconstructed all urban elements, such as benches and lampposts, as well as other environmental elements, including nearby buildings, trees, and the sky.

\subsection{Hardware}
\begin{figure*}
    \centering
  \includegraphics[width=.9\textwidth]{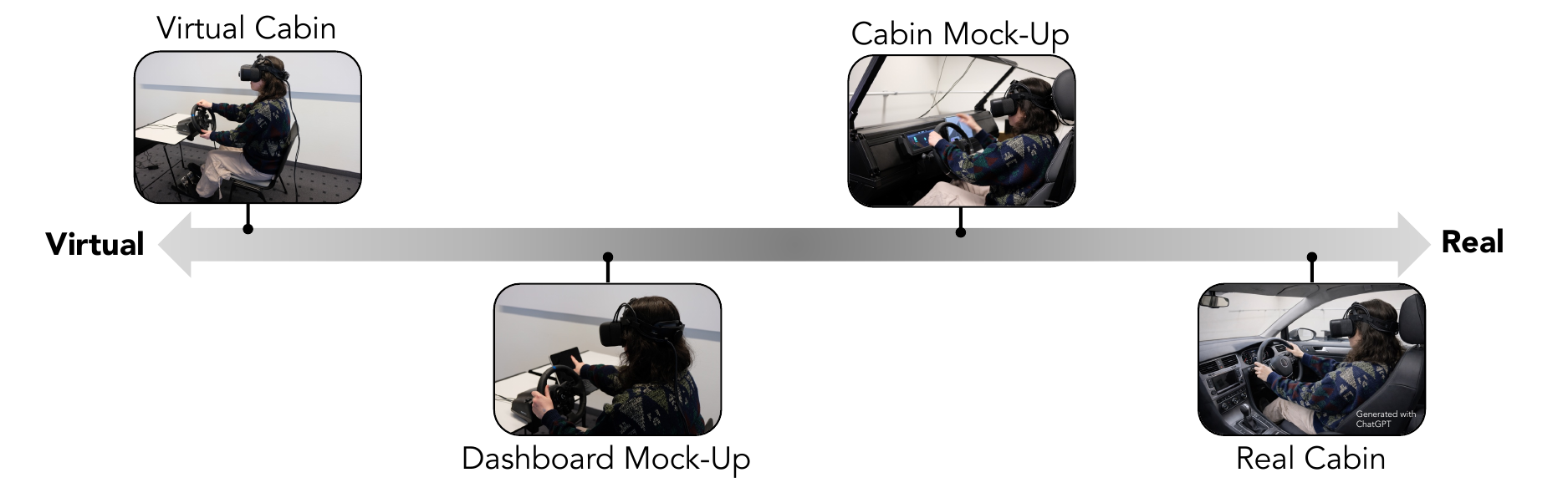}
  \caption{\name supports different levels of mock-up integration, ranging from a fully virtual cabin to a real vehicle.}
  \Description{The image shows a time scale ranging from virtual (left) to real (right), indicating the level }
  \label{fig:fidelity-scale}
\end{figure*}

\begin{figure*}
    \centering
  \includegraphics[width=\textwidth]{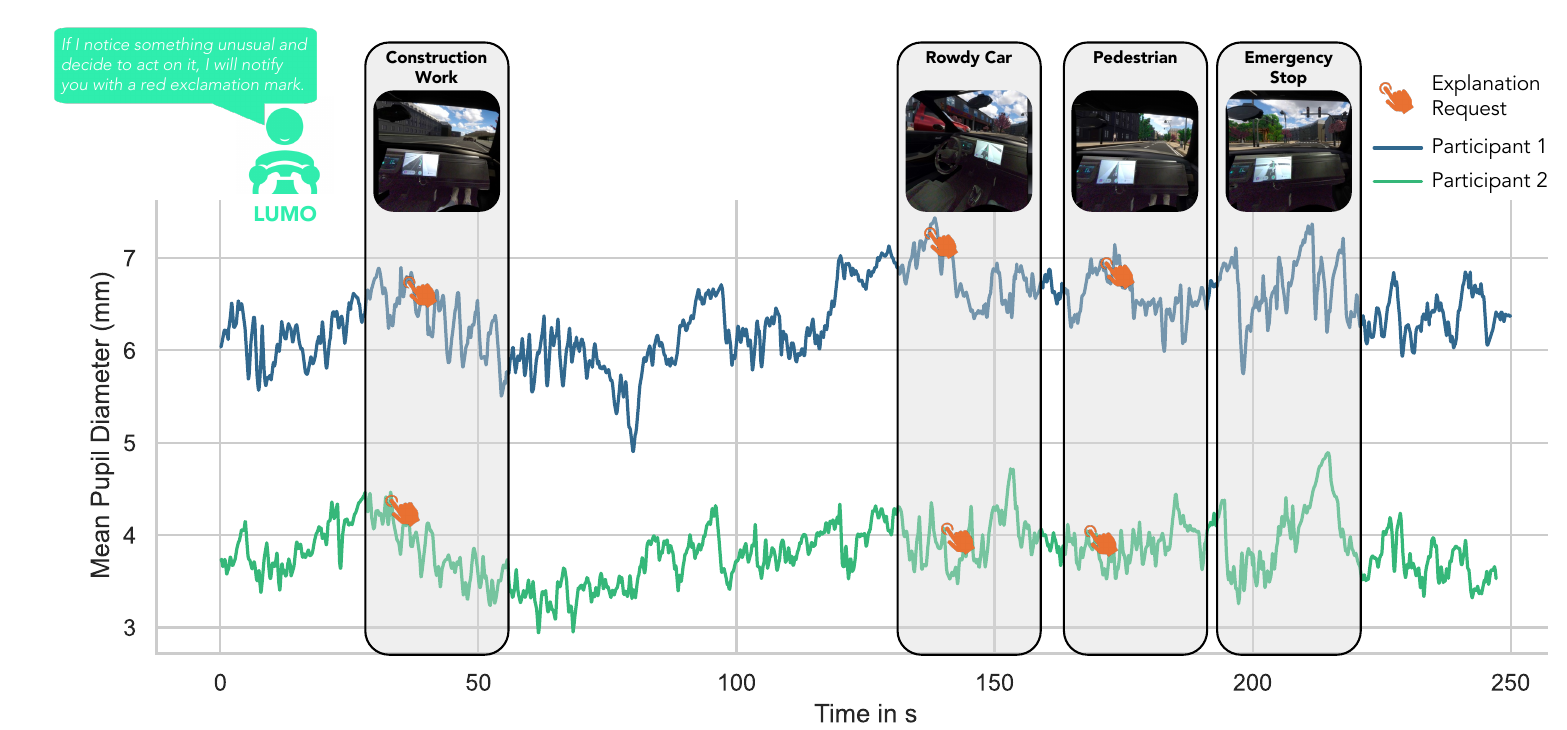}
  \caption{This graph shows how the pupil diameter of the two study participants changed over the course of the experiment and when participants decided to request an explanation by interacting with the center stack touchscreen.}
  \Description{TBD}
  \label{fig:results}
\end{figure*}

\paragraph{\textbf{Cabin Mock-Up}}

\name supports cabin mock-ups at different levels of fidelity. As illustrated in \autoref{fig:fidelity-scale}, \name can be used as a fully virtual driving simulator in which the entire vehicle interior is virtually rendered without any video pass-through. At the other extreme, \name can be deployed with a real vehicle. Between these two ends of the spectrum, a wide range of hybrid configurations is possible, such as setups that include only a physical dashboard or steering wheel, as well as high-fidelity cabin mock-ups that replicate most of the vehicle interior. This flexibility allows researchers to choose the level of physical realism that best fits their experimental goals. All setups that integrate any real-world parts require a 1:1 scale 3D software model of the vehicle interior. This model must be spatially aligned in the simulator software with the physical mock-up and rendered transparent so that the real-world components are visible in their correct virtual positions.

In this setup, we combine \name with a high-fidelity cabin mock-up from Ergoneers. The mock-up includes two electrically adjustable car seats, an OEM steering wheel coupled to a force-feedback motor, brake and accelerator pedals, an instrument cluster display, and a 17-inch center-stack touchscreen. This configuration provides a realistic in-car experience while remaining compact enough to keep space requirements manageable.

\paragraph{\textbf{Head-Mounted Display}}

\name requires a head-mounted display that combines VR and video pass-through capabilities into one device. Furthermore, it must allow editing of the pass-through objects in shape, size, and position to adjust the simulation and real-world cockpit to the mockup-setup. 
Therefore, we use Varjo XR-3 glasses, which provide a Unity library that allows precise adjustments to these relevant pass-through factors.

\paragraph{\textbf{Additional Hardware}}
To track the exact position and orientation of the \gls{HMD} \name requires Lighthouses to be positioned according to the requirements defined by the \gls{HMD} manufacturer. In the specific case, we use four \textit{SteamVR Base Stations 2.0}.
To align the 3D cabin model with the physical mock-up in position and rotation \name uses external trackers (i.e., the \textit{HTC Vive Tracker}), which should be mounted at a fixed location on the cabin mock-up. Compared to marker-based tracking or hard-coded positions, this approach provides more stable alignment and allows the entire mock-up to be freely repositioned within the tracking volume.

\section{Pilot Study}
To illustrate the capabilities of MRDrive, we conducted a small (N=2) pilot study. The purpose of this study was not to test behavioral hypotheses, but to demonstrate how MRDrive can be used to conduct multimodal user studies involving eye-tracking and touch interaction in a realistic driving context. Accordingly, we prototyped three in-vehicle agents and explored how multimodal explanations affect driver workload, measured through pupil dilation~\cite{pomplun2019pupil}.

\paragraph{\textbf{Study Procedure and Data Collection}}
Participants experienced several short automated driving scenarios as passengers and interacted with an in-vehicle interface that explained vehicle behavior. They were seated in the passenger seat of a mock-up vehicle that included a center-stack touchscreen featuring an anthropomorphic assistant that communicated driving explanations. They wore a Varjo XR-3 mixed-reality headset with integrated eye tracking. Each driving scenario included 4 safety-critical events for which, depending on the condition, explanations were available.
Each participant experiences three routes under three different intelligent drivers (Nelo (no explanations), Coda (proactive explanations), and Lumo (on-demand explanations)). Explanations were delivered via a speech and text displayed on the center stack touchscreen.
To compare the workload induced by the three different intelligent drivers and the rate at which participants requested explanations in the on-demand condition, we collected eye-tracking data (position data and pupil dilation) and touchscreen interactions.

\paragraph{\textbf{Results}}

\autoref{fig:results} shows the mean pupil diameter for both participants together with their touch-based explanation requests in the on-demand condition (Lumo). For both participants, pupil dilation tends to increase during potentially safety-critical events for which the intelligent driver provides explanations. While explanations were requested in the first three scenarios, no requests were made during the final emergency-stop scenario. Overall, these results illustrate how \name can be used to conduct automotive user research studies.

\section{Conclusion and Future Work}

In this work, we present \name, an open-source \gls{MR} driving simulator and flexible, high-fidelity platform for prototyping and evaluating \gls{IVIS} and driver interactions. By adopting \gls{MR}, \name combines the immersive 360\degree virtual driving environment with a real-world cabin mock-up, preserving haptic feedback and embodied interaction while maintaining experimental control over traffic scenarios and interface designs.
The \gls{MR} paradigm offers a middle ground between \gls{VR} simulators (which sacrifice physical realism) and expensive high-end projection-based setups. Although it entails higher setup complexity and hardware demands than basic \gls{VR} configurations, its modular open-source architecture built on Unity lowers the barrier for adoption.
Through a pilot study, we demonstrated how \name supports the collection and analysis of eye-tracking and touch interaction data. We see \name as a reusable research instrument that reduces the barrier to studying attention, explainability, and interaction in realistic driving scenarios. 
However, we acknowledge that certain limitations remain. Although the Varjo XR-3 is comparatively inexpensive compared to projection-based systems, it is a high-end headset that is required to align real and virtual content with a precision of <1\,cm. Additionally, alignment still requires a considerable amount of trial and error. Furthermore, in order to achieve frame rates of 40–60 fps, we use a workstation equipped with a high-end NVIDIA GeForce RTX 4090 graphics card. 
In addition to technical improvements such as more efficient rendering and even better alignment, future work could involve expanding the range of supported interaction modalities (e.g., gesture and voice input), integrating additional physiological sensors, and conducting further validation of the simulator against real-world driving benchmarks. This would strengthen its role in advancing human-centred design for future (automated) mobility. To support the open science movement in HCI~\cite{salehzadeh_niksirat_changes_2023, ebel_changing_2024}, we make \name openly available at: \url{https://github.com/ciao-group/mrdrive}

\bibliographystyle{ACM-Reference-Format}
\bibliography{references}
\end{document}

%% file: acronyms.tex
% \begin{acronym}[]
% 	\acro{AI}{Artificial Intelligence}
% 	\acro{ACC}{Adaptive Cruise Control}
% 	\acro{ADAS}{Advanced Driver Assistance System}
%     \acro{AV}{Automated Vehicle}
% 	\acro{CAN}{Controller Area Network}
% 	\acro{ECU}{Electronic Control Unit}
% 	\acro{FNN}{Feedforward Neural Network}
% 	\acro{GDPR}{General Data Protection Regulation}
% 	\acro{GOMS}{Goals, Operators, Methods, Selection rules}
% 	\acro{HCI}{Human-Computer Interaction}
% 	\acro{HMI}{Human-Machine Interface}
%     \acro{HMD}{Head-Mounted Display}
% 	\acro{HU}{Head Unit}
%     \acro{IQR}{Inter Quartile Range}
% 	\acro{IVIS}{In-Vehicle Information System}
% 	\acro{KLM}{Keystroke-Level Model}
% 	\acro{KPI}{Key Performance Indicator}
% 	\acro{MGD}{Mean Glance Duration}
%     \acro{MAE}{Mean Absolute Error}
%     \acro{ML}{Machine Learning}
%     \acro{MR}{Mixed Reality}
% 	\acro{OEM}{Original Equipment Manufacturer}
% 	\acro{OTA}{Over-The-Air}
% 	\acro{AOI}{Area of Interest}
%     \acro{TOR}{Take-over request}
% 	\acro{RF}{Random Forest}
% 	\acro{SA}{Situation Awareness}
%     \acro{SEEV}{Salience, Effort, Expectancy, and Value}
% 	\acro{SHAP}{SHapley Additive exPlanation}
% 	\acro{TGD}{Total Glance Duration}
% 	\acro{UCD}{User-centered Design}
% 	\acro{UX}{User Experience}
%     \acro{VR}{Virtual Reality}
% 	\acro{XAI}{Explainable AI}
%         \acro{IVIS}{In-Vehicle Information Systems}
%     \acro{MR}{Mixed Reality}
%     \acro{VR}{Virtual Reality}
%     \acro{XR}{Extended Reality}
% \end{acronym}

\newacronym{AI}{AI}{Artificial Intelligence}
\newacronym{ACC}{ACC}{Adaptive Cruise Control}
\newacronym{ADAS}{ADAS}{Advanced Driver Assistance System}
\newacronym{AV}{AV}{Automated Vehicle}
\newacronym{CAN}{CAN}{Controller Area Network}
\newacronym{ECU}{ECU}{Electronic Control Unit}
\newacronym{FNN}{FNN}{Feedforward Neural Network}
\newacronym{GDPR}{GDPR}{General Data Protection Regulation}
\newacronym{GOMS}{GOMS}{Goals, Operators, Methods, Selection Rules}
\newacronym{HCI}{HCI}{Human-Computer Interaction}
\newacronym{HMI}{HMI}{Human-Machine Interface}
\newacronym{HMD}{HMD}{Head-Mounted Display}
\newacronym{HU}{HU}{Head Unit}
\newacronym{IQR}{IQR}{Inter Quartile Range}
\newacronym{IVIS}{IVIS}{In-Vehicle Information System}
\newacronym{KLM}{KLM}{Keystroke-Level Model}
\newacronym{KPI}{KPI}{Key Performance Indicator}
\newacronym{MGD}{MGD}{Mean Glance Duration}
\newacronym{MAE}{MAE}{Mean Absolute Error}
\newacronym{ML}{ML}{Machine Learning}
\newacronym{MR}{MR}{Mixed Reality}
\newacronym{OEM}{OEM}{Original Equipment Manufacturer}
\newacronym{OTA}{OTA}{Over-The-Air}
\newacronym{AOI}{AOI}{Area of Interest}
\newacronym{TOR}{TOR}{Take-over Request}
\newacronym{RF}{RF}{Random Forest}
\newacronym{SA}{SA}{Situation Awareness}
\newacronym{SEEV}{SEEV}{Salience, Effort, Expectancy, and Value}
\newacronym{SHAP}{SHAP}{SHapley Additive exPlanation}
\newacronym{TGD}{TGD}{Total Glance Duration}
\newacronym{UCD}{UCD}{User-Centered Design}
\newacronym{UX}{UX}{User Experience}
\newacronym{VR}{VR}{Virtual Reality}
\newacronym{XAI}{XAI}{Explainable AI}
\newacronym{XR}{XR}{Extended Reality}